\documentclass[prl,aps,twocolumn,groupaddress,longbibliography]{revtex4-1}
\usepackage{latexsym}
\usepackage{lineno}
\usepackage{hyperref}
\usepackage{url}
\usepackage{graphicx}
\usepackage{verbatim}
\usepackage{multirow}
\usepackage{amsmath}
\newcommand{\beq}{\begin{eqnarray}}
\newcommand{\eeq}{\end{eqnarray}}
\usepackage{mathrsfs}
\usepackage{float}
\usepackage[usenames, dvipsnames]{color}
\usepackage{mathtools}
\usepackage{slashed}
\usepackage{physics}	
\usepackage{graphicx}   
\usepackage{epstopdf}
\usepackage{subfigure}  
\usepackage{hyperref}   
\usepackage{bbold}
\usepackage{wasysym}
\usepackage{amssymb}
\usepackage{feynmp}
\usepackage[symbol]{footmisc}
\def \bs{\textbf}
\usepackage[latin1]{inputenc}
\usepackage{tikz}
\usetikzlibrary{decorations.pathmorphing}
\usetikzlibrary{shapes.misc}
\tikzset{cross/.style={cross out, draw=black, minimum size=8*(#1-\pgflinewidth), inner sep=0pt, outer sep=0pt},
cross/.default={1pt}}
\usetikzlibrary{patterns,math}
\newcommand{\RN}[1]{%
  \textup{\uppercase\expandafter{\romannumeral#1}}%
}
\begin{document}

\title{Superconductivity from Luttinger surfaces: Emergent $\infty$-body SYK physics}
\author{Chandan Setty}
\thanks{email for correspondence: csetty@rice.edu}
\affiliation{Department of Physics, University of Florida, Gainesville, Florida, USA}
\affiliation{Department of Physics and Astronomy, Rice University, Houston, Texas, USA}

\begin{abstract}
The pairing of two electrons on a Fermi surface due to an infinitesimal attraction between them always results in a superconducting instability at zero temperature ($T=0$). The equivalent question of pairing instability on a Luttinger surface (LS) -- a contour of zeros of the propagator -- instead leads to a quantum critical point (QCP) that separates a non-Fermi liquid (NFL) and superconductor. A surprising and little understood aspect of pair fluctuations at this QCP is that their thermodynamics maps to that of the Sachdev-Ye-Kitaev (SYK) model in the strong coupling limit. Here, we offer a simple justification for this mapping by demonstrating that (i) LS models share the reparametrization symmetry of the $q\rightarrow \infty$ SYK model with $q$-body interactions \textcolor{black}{close to the LS}, and (ii) the enforcement of gauge invariance results in a $\frac{1}{\sqrt{\tau}}$ ($\tau\sim T^{-1}$) behavior of the fluctuation propagator near the QCP, as is a feature of the fundamental SYK fermion.

\end{abstract}
\maketitle
\section{Introduction} 
The theory of superconductivity by Bardeen, Cooper and Schrieffer~\cite{BCS1957} relies on the existence of a Fermi surface (FS)-- a contour of \textit{poles} of the single particle Green function where excitations are long-lived and the notion of a quasiparticle is well-defined.  The superconducting phase then follows from a net attractive interaction that pairs two such quasiparticles rendering the FS unstable.
 The question of whether such an instability can exist, or for that matter be defined, on a Luttinger surface (LS)~\cite{AGD1965}, or a contour of \textit{zeros} of the propagator due to a divergent self-energy~\cite{AGD1965, Campuzano1998, Tsvelik2002, Tsvelik2006, YRZ2006, Kotliar2006, Hong-Phillips2012, Sachdev2018}, is less straightforward.  The obstacle to such a generalization stems from a complete breakdown of the quasiparticle concept on a LS -- the quasiparticle scattering lifetime is vanishingly small in this limit and the degrees of freedom that constitute a ``Cooper pair" are either ill-defined or unknown due to lack of exactly solvable models. 
Yet, experiments suggest umpteen examples of pairing in highly incoherent matter~\cite{Taillefer2010}, including the Cuprates where LSs play a prominent role in their normal state phenomenology~\cite{Campuzano1998, Statt1999, Uchida2006, Shen2010,Shen2011, Kidd2011, Chakravarty2010, Dzyaloshinskii1996, Dzyaloshinskii2003, Tsvelik2002, KRT2006, YRZ2006, Morr1998, Kotliar2006, Rosch2007, Kane2013, Georges2006, Imada2009, Phillips2006, Bascones2007, Campuzano2007, Vanacore2018, Choy2007, Sachdev2018}\par  
In a recent attempt~\cite{Setty2020} to address the question above, it was demonstrated phenomenologically that such a pairing instability can indeed exist on a minimal model LS where the self-energy has a simple pole. Unlike the case of a Fermi liquid, 
the pair susceptibility in this scenario diverges even at zero temperature resulting in a superconductor to non-Fermi liquid (NFL) quantum phase transition above a critical residue of the self-energy pole. In addition to a power-law divergence of the spectral density, a surprising new feature of the NFL phase close to the quantum critical point (QCP) was uncovered -- the pair fluctuation free energy resembles that of the Sachdev-Ye-Kitaev (SYK)~\cite{SY1993, Kitaev2015, Stanford2016} model in the limit where the self-energy residue is much greater than the temperature (strong coupling limit). The SYK model has garnered recent interest (see~\cite{Rosenhaus2019} and references therein) due to its connections to gravitational models describing black holes~\cite{Kitaev2015}. Motivated by these observations, Phillips and coworkers~\cite{Phillips2019} revived an exactly solvable but startlingly simple microscopic model by Hatsugai and Kohmoto (HK)~\cite{HK1992} that hosts \textcolor{black}{finite frequency} LSs in the Mott phase. Upon doping the Mott insulator, 
they found that the elementary pair excitations are formed by \textit{doublons} and \textit{holons} as opposed to Bogoliubov quasiparticles. That such a mapping between the SYK model and LS fluctuation thermodynamics should work so well, yielding an exact but simple picture of pair excitations upon doping, is unsettling and can be critiqued as coincidental -- so far, there exists no symmetry based argument invoked whatsoever to justify the mapping, nor is there a consistent analysis that places interaction vertices and self-energies on an equal footing in accordance with the Ward-Takahashi identity~\cite{Nambu1960, Schrieffer2018}.  \par 
It is the purpose of this paper to show that LS models, \textcolor{black}{such as the one by HK, share the symmetries of the $q\rightarrow \infty$ limit of the SYK model with a $q$-body interaction}, and that the propagator resulting from gauge-invariant pair fluctuations on a LS is the SYK conformal Green function. Thus our work gives a simple justification for the robustness of the mapping between pair fluctuations on a LS and SYK models. More specifically, we demonstrate that LS models where the self-energy has a simple pole acquire an infinite-$q$ reparametrization symmetry in the low-energy scaling limit.  The proof we provide follows along the lines of its original formulation in context of the SYK model by the authors of Refs.~\cite{Sachdev2015, Kitaev2015, Stanford2016}.  To contrast, however, we work with a self-energy that is proportional to the single particle \textit{non-interacting} Green function -- a characteristic of the HK or phenomenological models proposed in Refs.~\cite{YRZ2006, KRT2006} where the original FS is converted into a LS.  This is a key feature of our analysis that distinguishes it from the random matrix model (or the $q=2$ SYK model). In the latter scenario, the self energy is proportional to a linear power of the total Green function in the presence of random hopping matrix elements; therefore, there exist no propagator zeros rendering the model effectively non-interacting. \par
 A further exact gauge-invariant evaluation of the pair response from the model LS Green function is essential for the theory to maintain charge conservation. This treatment follows from the Ward-Takahashi~\cite{Nambu1960, Schrieffer2018} identity and ensures that interaction vertices and self-energies are placed on an equal footing. Recent progress has been made in this direction for several phenomenological models describing the Cuprate pseudo-gap~\cite{Levin2016, Boyack2017, Lee2017, Guo2017, He2018}. We find that gauge-invariance results in a $\frac{1}{\sqrt{\tau}}$ ($\tau\sim T^{-1}$) behavior of the fluctuation propagator in the strong coupling limit -- a feature of the fundamental, particle-hole asymmetric, SYK fermion. As a consequence, the fluctuation free-energy, computed exactly from the gauge-invariant pair response, mimics leading order SYK fluctuations in the strong coupling limit. Furthermore, we find that gauge-invariance enforces the fluctuation density of states obtained from the partition function in the static, long-wavelength limit to exhibit a $\omega^{-1}$ behavior at low energies indicating small energy spacing in the fluctuation spectrum. In the weak coupling limit, vertex corrections only have quantitative effects on the phase diagram.  The conclusions of our work point toward the existence of a mapping between effective theories of fluctuations on model LSs such as the HK model and models of gravity.     \par
 \section{Models and Results} 
\textit{Models:} Various phenomenological~\cite{Campuzano1998,YRZ2006, KRT2006}, microscopic~\cite{Baskaran1991, HK1992, Tsvelik2002, Tsvelik2006, Kotliar2006,Rosch2007, Imada2009, Ohta2011, Hong-Phillips2012, Kane2013, Sachdev2018}  and holographic~\cite{Phillips-Leigh2011-1, Phillips-Leigh2011-2}  models have been used as relevant starting points for describing LSs. Some of these models have been used extensively to understand and interpret
 ~\cite{ Dzyaloshinskii1996, Dzyaloshinskii2003, Tsvelik2002, Tsvelik2006, YRZ2006, Morr1998, Kotliar2006, Choy2007, Rosch2007, Kane2013, Georges2006, Imada2009, Phillips2006, Bascones2007, Campuzano2007, Vanacore2018,  Sachdev2018} a variety of experimental observations in the strongly interacting phase of the Cuprates~\cite{Statt1999, Uchida2006, Shen2010,Shen2011, Kidd2011, Chakravarty2010}. The simplest among them are models where the full interacting inverse Green function,  $G(\bs p, i\epsilon_n)^{-1} = i \epsilon_n - \xi(\bs p) - \Sigma(\bs p, i\epsilon_n) $,  contains a self-energy of the form~\cite{Campuzano1998,YRZ2006, KRT2006}
 \beq
\Sigma(\bs p, i \epsilon_n) =  \frac{u^2}{i \epsilon_n + \xi(\bs p)}.
\label{SelfEnergy}
\eeq
Here we have defined $\xi(\bs p) = \epsilon(\bs p) - \mu$ as the band dispersion with chemical potential $\mu$, $\epsilon_n$ is the fermionic frequency and $u$ sets the energy scale of the residue of the pole. The microscopic HK Hamiltonian offers an interpretation of the parameter $u$ as a four-body interaction term where only scattering process which conserve the total center of mass are included. In Fourier transformed space, this Hamiltonian takes a particularly simple form 
\begin{equation}
	H_{\rm HK} = \sum_{\bs k}H_{\bs k} = \sum_{\bs k} \left[\xi_{\bs k}(n_{\bs k\uparrow} + n_{\bs k\downarrow} ) + 2 u ~n_{\bs k\uparrow} \, n_{\bs k\downarrow}\right],
\label{kSpaceHK}
\end{equation}
where $n_{\bs k\sigma}$ is the number operator for a state with momentum $\bs k$ and spin $\sigma$. Hence, the HK model is simply a Hubbard model in momentum space where the interaction term commutes with the kinetic energy and is exactly solvable. In the scenario where the interaction $2 u$ is larger than the bandwidth of the non-interacting bands, there exists a Mott gap between an upper and lower Hubbard-like bands (although the model is too simple to capture any dynamical spectral weight transfer effects which the Hubbard model does). \textcolor{black}{ A more recent analysis~\cite{Yang2020} of Fermi arcs and pseudo-gap in the HK model lead to the interpretation of $u$ as the pseudo-gap order parameter in the normal state. This is in line with expectation from previous works~\cite{Campuzano1998, KRT2006, YRZ2006}}. The propagator of the model Hamiltonian in Eq.~\ref{kSpaceHK} depends on the occupation numbers of the upper and lower Hubbard bands, and a LS is not generally obtained at all momenta for arbitrary chemical potential and occupation numbers. However, when the occupancies of the upper and lower bands are equal and the chemical potential lies between the bands,
 a LS with self-energy written in Eq.~\ref{SelfEnergy} with a (renormalized) dispersion $\xi(\bs p) \rightarrow -\xi(\bs p)$ is ensured at all momenta \textcolor{black}{ (see Refs.~\cite{HK1992, Setty2018, Phillips2019} for the self-energy in the HK model). This statement is exact and non-perturbative and holds even when the parameter $u$ is taken to infinity.} Therefore, close to their respective LSs, the distinction between the Green function of Eq.~\ref{kSpaceHK} and that corresponding to model Eq.~\ref{SelfEnergy} vanishes. 
 \par 
\textit{Reparametrization symmetry:} We now take a closer look at the self-energy (Eq.~\ref{SelfEnergy}) appearing in the total Green function $G(\bs p, i\epsilon_n)$. This equation can be rewritten as 
\textcolor{black}{
\beq
\Sigma(\bs p, i \epsilon_n) =  -u^2 G_0(\bs p, -i\epsilon_n)
\label{Repara1}
\eeq
where $G_0(\bs p, i\epsilon_n)$ is the non-interacting Green function.} 
 \textcolor{black}{As in the zero dimensional SYK model~\cite{Kitaev2015, Stanford2016}, we will be interested in the low-energy scaling limit $i\epsilon_n \rightarrow 0$ where one anticipates reparametrization invariance for finite interactions. Equivalently, this corresponds to a Green function contribution on the LS ($\xi(\bs p)=0$) where the self-energy diverges and its spatial structure is washed out as $\beta u \rightarrow \infty$. Therefore, the following equations hold in this limit}
\beq
G(\bs p, i\epsilon_n) &\simeq& -\Sigma(\bs p, i\epsilon_n)^{-1} \simeq -\frac{i\epsilon_n}{u^2} \equiv G(i\epsilon_n)
\label{Repara22}
\\
\Sigma(\bs p, i\epsilon_n) &\simeq& \frac{u^2}{i\epsilon_n} \equiv u^2 G_0(i\epsilon_n) \equiv \Sigma(i\epsilon_n).\label{Repara21} 
\eeq
Here we have defined the local quantities $G(i\epsilon_n)$, $G_0(i\epsilon_n)$,  and $\Sigma(i\epsilon_n)$. Transforming into imaginary time coordinate and invoking time translational invariance we have the total Green function from Eq.~\ref{Repara22}
\beq
G(\tau - \tau') &=& \frac{\delta'(\tau-\tau')}{u^2}
\label{TauSpaceG}
\eeq
where $\delta'(\tau)$ is the temporal derivative of the Dirac delta function. We can also evaluate the imaginary time self-energy as well as the convolution integral of $G(\tau-\tau') \Sigma(\tau'-\tau'')$ in the same limit. Using Eqs.~\ref{Repara1},~\ref{Repara22},~\ref{Repara21}, this gives us 
\beq\label{Invariant1}
\Sigma(\tau-\tau') = u^2 G_0(\tau-\tau') &=&  \frac{u^2}{2} \text{sgn}(\tau - \tau') \\
\int d\tau'G(\tau-\tau') \Sigma(\tau'-\tau'') &=& -\delta(\tau-\tau''). \label{Invariant2}
\eeq
The equations~\ref{Invariant1} and ~\ref{Invariant2} above are written in proximity to the LS and have some notable features. First, as in SYK type models, they are invariant under a reparametrization symmetry transformation $\tau \rightarrow f(\tau)$ and
\beq\label{Scaling1}
G(\tau - \tau') &\rightarrow& \left[f'(\tau') f'(\tau)\right] G\left(f(\tau) - f(\tau')\right)\\
\Sigma(\tau - \tau') &\rightarrow& \Sigma\left( f(\tau) - f(\tau')\right).
\label{Scaling2}
\eeq

\begin{figure}
\includegraphics[width=3.2in,height=0.6in]{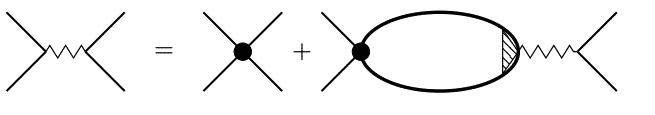} \hfill
\caption{Fluctuation propagator in the Cooper channel (zig-zag lines) defined through the Bethe-Salpeter equations. The black solid disk and the shaded triangle denote the bare interaction vertex and vertex corrections arising from electron correlations respectively. The thick black lines denote total electron Green function.} \label{Feynman}
\end{figure}
The above scaling behavior of $G(\tau - \tau')$ and $\Sigma(\tau - \tau')$ in the Eqs.~\ref{Scaling1} and~\ref{Scaling2} under reparametrization is same as the $\Delta^{-1}\equiv q\rightarrow \infty$ limit of the $q$-body SYK model~\cite{Stanford2016} \textit{provided} the roles of $G(\tau - \tau')$ and $\Sigma(\tau - \tau')$ are swapped (the other choice of $\Delta^{-1}\equiv q = 1$ is unphysical). This exchange is an important feature that completes the map to the conformal limit of infinite-body SYK model. More specifically, it highlights the duality between  \textcolor{black}{ self-energies in gapped models (of the form appearing in Eq.~\ref{SelfEnergy}) close to a LS and propagators in CFTs. That is, a theory  characterized by a propagator defined by the self-energy of a gapped LS is conformal in the sense of the derivation leading to Eq~\ref{Scaling2}.}    
\textcolor{black}{ 
 It is notable that in the special limit of $q\rightarrow \infty$, the SYK model also has a reparametrization invariant propagator; however, the propagator in Eq~\ref{Scaling1} only transforms covariantly. Similarly, the self-energy in Eq.~\ref{Scaling2} maintains the full symmetry of the reparametrization group. This is unlike the SYK model where the self-energy transforms only covariantly under reparametrizations for all $q\geq 2$. In both cases, however, the transformation properties of the propagator and self-energies act to leave Eqs.~\ref{Invariant1} and ~\ref{Invariant2} reparametrization invariant.} This is the key reason why the generic form of fluctuations of the two aforementioned duals cannot be distinguished from one another (as will be shown below).  
Second, the forms of the Eqs.~\ref{Invariant1} and ~\ref{Invariant2}, while seemingly similar to the random matrix model (or the $q=2$ SYK model), are not the same -- in the current model with a LS, the self-energy in Eq.~\ref{Invariant1} is proportional to the linear power of the \textit{non-interacting} Green function as opposed to the random matrix model where the full Green function appears. This is an important distinction as the solution of the random matrix model does not contain zeros of the Green function and is therefore an effectively non-interacting in the presence of long-range disorder. \textcolor{black}{Finally}, it can be verified that arguments leading to Eqs.~\ref{Invariant1}, \ref{Invariant2}, \ref{Scaling1} and \ref{Scaling2} also hold for the diagonal elements of a BCS Green function with the interaction parameter $u$ replaced by the superconducting order parameter. In this sense, they belong to the same reparametrization symmetry class near the LS. However, our focus in the following section is on fluctuations in the non-superconducting state; hence, we will not have off-diagonal long range order through anomalous contributions to the pair susceptibility. \textcolor{black}{At this point, we emphasize that while the parameter $u$ begs such an interpretation, it does not play the role of a  superconducting gap in our work. It is rather the interaction (``Mott") gap that exists even above the instability temperature. 
 From Eqs.~\ref{Invariant1}~\ref{Invariant2} and ~\ref{Scaling1}~\ref{Scaling2}, and the remarks that follow, we rightly anticipate -- further supported by} the original derivation for the SYK model~\cite{Stanford2016} and later for LS models with a simple pole in the self-energy~\cite{Setty2020} -- that the leading order free energy contribution from pair-fluctuations in the NFL phase close to the QCP takes the form $-\beta F = \beta u^* - \gamma \ln(\beta u^*)$, where $u^*$ is the QCP, $\beta$ is inverse temperature, and $\gamma$ is a constant that determines the fluctuation density of states. This is demonstrated below.\par 
 \textit{Ward-Takahashi identity and exact vertex:} We now set out to evaluate the fluctuation free energy and density of states from the gauge-invariant fluctuation propagator and pair susceptibility by approaching the QCP from the NFL side.  The Bethe-Salpeter equation for the fluctuation propagator is shown in Fig.~\ref{Feynman} in terms of the fully interacting pair bubble. An immediate difficulty in evaluating the gauge-invariant pair bubble is that one requires a knowledge of the exact fluctuation vertex in models of LSs. This is a-priori not straightforward if one chooses to obtain the vertex directly from the Hamiltonian, especially in LS ansatz models such as those espoused in Refs.~\cite{YRZ2006, KRT2006} where the underlying electronic degrees of freedom are unspecified. However, and regardless of whether the initial Hamiltonian is known, one can resort to the Ward-Takahashi identity~\cite{Nambu1960, Schrieffer2018} to obtain the full vertex \textit{provided} the exact self-energy is known. This is because once the self-energy is fixed, charge conservation restricts the form of the vertex function. In fact this approach has been recently advocated in several works describing phenomenological models of the Cuprate pseudo-gap~\cite{Levin2016, Boyack2017, Lee2017, Guo2017, He2018}. \par
\begin{figure}
\includegraphics[width=3.2in,height=0.5in]{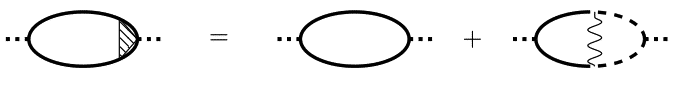} \hfill
\caption{Feynman diagrams defining the vertex corrections through the Ward identity. The thick solid (dashed) lines are the interacting (non-interacting) Green functions. The shaded triangle denotes vertex corrections arising from electron correlations. The dotted lines denote a generic external momentum transfer into and away from the pair bubble.} \label{Feynman2}
\end{figure}
Since the self-energy is known exactly in our case, we can proceed with the Ward-Takahashi identity. This identity relates the pair-vertex to the interacting Green function. For a Matsubara frequency $i q_n$ and momentum transfer $\bs q$, it takes the form
\beq \nonumber
-i q_n \Gamma_0(p+q, p) + \bs q \cdot \mathbf{\Gamma} (p+q, p) 
= G^{-1}(p) - G^{-1}(p+q) \\
&&
\eeq
where the vertex function $\Gamma_{\mu} \equiv \left(\Gamma_0, \mathbf{\Gamma}\right)$ for the charge and current and we introduce the notation $q \equiv \left(iq_n, \bs q \right)$. Using the definition $G(p)^{-1} = i p_n - \xi(\bs p) - \Sigma(p) $, we arrive at an expression for the exact vertex
\beq \nonumber
\Gamma_{\mu}(p+q, p) &=& \gamma_{\mu}\left(1 + \frac{u^2}{\left(ip_n + iq_n + \xi_{\bs p + \bs q}\right)\left(ip_n + \xi_{\bs p}\right)}\right) \\ 
&=& \gamma_{\mu}\left(1 \pm u^2 G_0(-p-q) G_0(-p)\right),
\label{Vertex}
\eeq
where $\gamma_{\mu} = \left(1, \bs p + \frac{\bs q}{2}\right)$ is the non-interacting vertex and the $\pm$ hold for the charge and current vertices respectively. As we are interested in fluctuations in the non-superconducting state, we have ignored collective mode contributions to the vertex that are proportional to the superconducting order parameter~\cite{Levin2016, Guo2017}. We now use the charge vertex in Eq.~\ref{Vertex} to obtain the gauge-invariant pair susceptibility (shown diagrammatically in Fig.~\ref{Feynman2})
\beq \nonumber
  \Pi(\bs q, i q_n) &=& \frac{1}{\beta (2\pi)^d} \sum_{\epsilon_n} \int d^d\bs p~G(p+q)G(-p)\Gamma_0(-p, p+q)\\
  &&\\
  &\equiv&   \Pi_0(\bs q, iq_n) + \Pi_{\Gamma}(\bs q, iq_n).
\eeq
Here we have decomposed the total susceptibility into the bare and interaction corrected vertex terms as appearing on the right hand side in Fig.~\ref{Feynman2}. \textcolor{black}{Note that we have defined the pair susceptibility above as well as in Ref.~\cite{Setty2020} in a ``symmetric" scheme (see for example~\cite{Levin2005} and references therein) where both the Green functions are interacting. This scheme leads to the $T=0$ ground state wave function described in~\cite{Phillips2019}. It is also interesting to seek solutions in the ``asymmetric"~\cite{Levin2005} scheme (one interacting and the other non-interacting Green function), which will generally yield a different ground state, but we will not address this case here.} We now proceed to evaluate the momentum integral by making a similar decomposition
\beq
I(q) &=& \int d^d\bs p~G(p+q)G(-p)\Gamma_0(-p, p+q) \\ 
&\equiv& I_0(q) + I_{\Gamma}(q)
\eeq
with the definitions $I_0(q) = \int d^d\bs p~G(p+q)G(-p) $ and $I_{\Gamma}(q) = u^2 \int d^d\bs p~G(p+q)G_0(-p-q)G(-p)G_0(p) $. Below we solve for the total pair susceptibility in the strong coupling limit where the interaction $u\gg T$. The effects of vertex corrections on the weak coupling result ($u\ll T$) in Ref.~\cite{Setty2020} is only quantitative and henceforth we ignore this case. \par
\begin{figure}[h!]
\includegraphics[width=3in,height=3in]{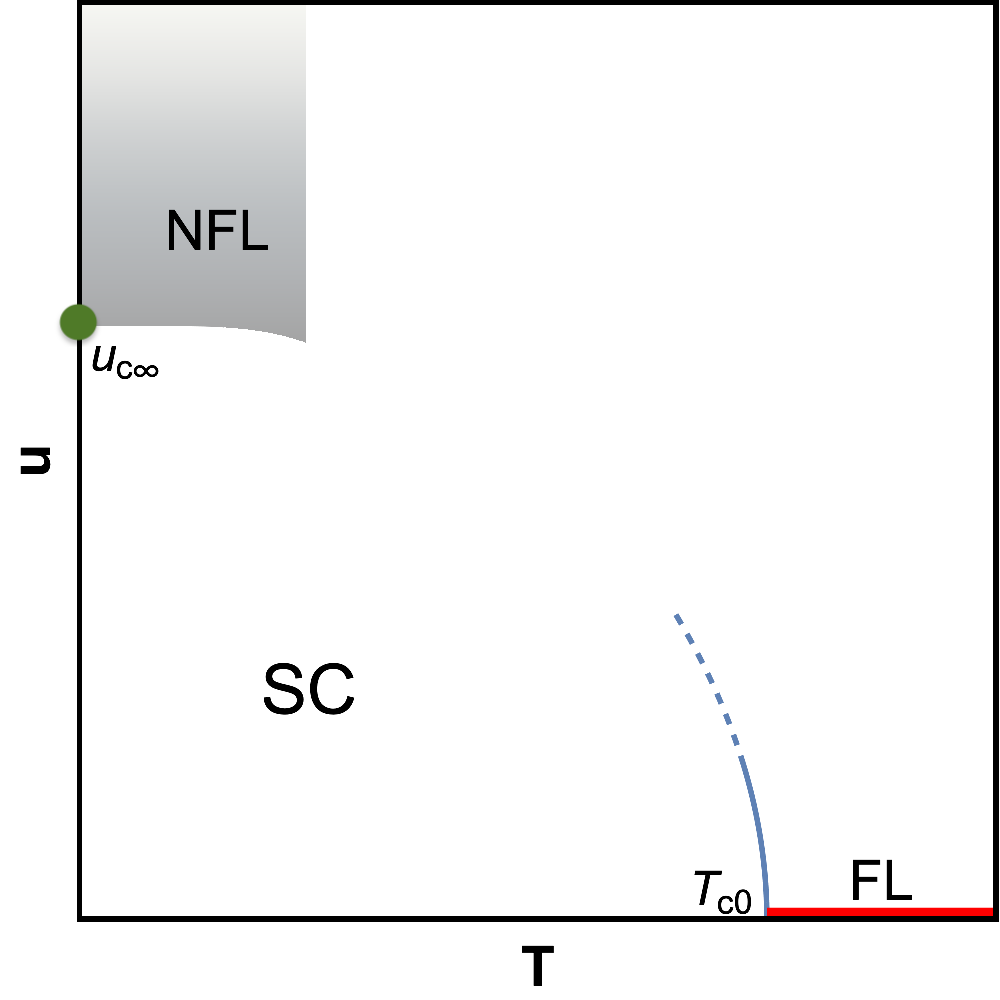} \hfill
\caption{ A schematic of the $u$-$T$ phase diagram. The gray (blue) contours are the strong coupling $\beta u \equiv \kappa \gg 1$ (weak coupling $\beta u \ll 1$) phase boundary and the red line denotes a Fermi liquid. The normal state at strong coupling is a non-Fermi liquid.  $T_{c0}$ is defined as $T_c(u=0)$ and $u_{c\infty}$ as $u_c(\beta\rightarrow \infty)$ (green dot).  }\label{PD}
\end{figure}
\textit{Strongly coupled ($\beta u \gg 1$) fluctuations:} In order to study the pairing instability, we are primarily interested in deriving the static, long-wavelength limit of the fluctuation propagator. Hence, we take the limiting conditions $iq_n \rightarrow 0$ and $r \equiv \frac{p_f |\bs q|}{m} \ll u$, where $m$ and $p_f$ are the mass and Fermi momentum from the non-interacting electronic dispersion respectively. In this regime, the expression for $\Pi_0(\bs q, iq_n) $ to second order in $r$ takes the form~\cite{Setty2020} (odd powers in $r$ vanish due to the angular integral)
\beq\label{BarePair}
\Pi_0(\bs q, iq_n\rightarrow 0) &\simeq&  \Pi^{(0)}_0(0,0) + \Pi^{(2)}_0(\bs q,0), \\ \nonumber
\Pi^{(0)}_0(0,0) &=& \frac{m}{4}\left(2 S_1 - u^2 S_3\right)\\ \nonumber
\Pi^{(2)}_0(\bs q,0) &=& -\frac{m r^2}{32}\left(2 S_3 - u^2 S_5\right),
\eeq
where  
 $S_{\nu} = \frac{1}{\beta} \sum_{\epsilon_n} (\epsilon_n^2 + u^2 )^{-\nu/2} $ with $\nu$ an odd integer. It should be noted that, for each order in $r$ that is non-vanishing, there is a term proportional to the residue $u^2$. Performing the small momentum expansion and using the same procedure described in Ref.~\cite{Setty2020}, on can similarly obtain the pair susceptibility from the vertex correction term given by $\Pi_{\Gamma}(\bs q, iq_n\rightarrow 0) \simeq  \Pi^{(0)}_{\Gamma}(0,0) + \Pi^{(2)}_{\Gamma}(\bs q,0)$ where
 \beq 
 \Pi^{(0)}_{\Gamma}(0,0) &=& \frac{m u^2}{4} S_3, \\ \nonumber
 \Pi^{(2)}_{\Gamma}(\bs q,0) &=&  -\frac{mu^2r^2}{32} S_5.
\eeq
In this case, however, only terms proportional to $u^2$ survive and have the opposite sign compared to those appearing in the bare susceptibility in Eqs.~\ref{BarePair}. Substituting the vertex correction terms back into the full expression for the pair susceptibility  $ \Pi(\bs q, iq_n) \equiv   \Pi_0(\bs q, iq_n) + \Pi_{\Gamma}(\bs q, iq_n)$, the terms proportional to $u^2$ from the bare susceptibility and vertex corrections cancel. 
The remaining Matsubara summations $S_{\nu}$ can be performed exactly, and we obtain the strong coupling ($\beta u \gg 1$), static, long-wavelength limit of the inverse fluctuation propagator,  denoted $L^{-1}(\bs q, iq_n\rightarrow 0)$, from Fig.~\ref{Feynman} as (see Ref.~\cite{Setty2020} for further details)
\beq \nonumber
L^{-1}(\bs q, iq_n\rightarrow 0) &=& - g^{-1} + N_0 \left[ \ln\frac{\Lambda}{u} - \frac{\sqrt{2\pi}~e^{-\beta u}}{\sqrt{\beta u}}\right] \\ 
&-& \frac{N_0 r^2}{8 u^2} \left( 1- \sqrt{2\pi \beta u}~e^{-\beta u}\right).
\label{FP}
\eeq
Here $g$ is the bare superconducting interaction, $N_0$ is the density of states of the non-interacting FS and $\Lambda$ is the Matsubara cut-off for the summation when $\nu=1$. We can now derive  the QCP separating the superconducting and NFL phases by setting $\beta u \rightarrow \infty$ and seeking a critical $u$  for which the fluctuation propagator diverges. This gives us $u^* \equiv u_{c\infty} = \Lambda e^{-1/N_0 g}$, and the contour of instability for low but non-zero temperatures is shown in the phase diagram in Fig.~\ref{PD}.  \textcolor{black}{A notable feature of the strongly coupled superconductor obtained here and in Ref.~\cite{Setty2020} is that it results from a gapped LS for a finite interaction $g$. That is, and unlike the BCS case, pairing in the current model is made possible by an attractive interaction $g$ that is above a critical value set by the formula $u = \Lambda e^{-1/N_0 g}$. 
Hence, doping does not play a crucial role in this problem.} This must be contrasted with quantum critical BCS superconductors obtained from pairing gapless conformal fermions~\cite{Zaanen2009, Zaanen2011} or mediated by quantum critical bosons~\cite{Chubukov2010}.
Fluctuation properties close to the QCP can be further extracted by setting $u=u_{c\infty}$ in the fluctuation propagator in Eq.~\ref{FP} which yields
\beq
L^{-1}\left(\bs q \rightarrow 0, iq_n =0 \right) |_{u\rightarrow u_{c\infty}} \simeq N_0 \frac{e^{-\beta u_{c\infty}}}{\sqrt{\beta u_{c\infty}}}.
\label{FPQCP}
\eeq
It is worthwhile to examine the consequences of Eq.~\ref{FPQCP} obtained by respecting the Ward identity. First, while gauge invariance leaves the QCP unaffected in comparison with the \textcolor{black}{non-gauge invariant} approximation~\cite{Setty2020}, it modifies the exponent of the $\beta u_{c\infty}$ factor in front of the exponential in Eq.~\ref{FPQCP} from $\frac{1}{2}$ to $-\frac{1}{2}$.  Therefore, the structure of the fluctuation propagator near zero temperature in Eq.~\ref{FPQCP} takes a form similar to the particle-hole asymmetric conformal propagator discussed in~\cite{Sachdev2019} with the substitution $T^{-1}\rightarrow \tau$. The $\frac{1}{\sqrt{\tau}}$ dependence is a signature of NFL transport and has been used to describe~\cite{Sachdev2019, Sachdev2015} Planckian behavior~\cite{Proust2019, Paglione2019} and universal scattering rates~\cite{Mackenzie2013} observed in a variety of strongly correlated materials. Second, following the methods described in~\cite{LarkinVarlamov2005, Glatz2018}, one can evaluate the fluctuation free energy from Eq.~\ref{FPQCP} to obtain $-\beta F = \beta u_{c\infty} - \gamma \ln(\beta u_{c\infty})$ with $\gamma = -\frac{1}{2}$. Thus, despite a change in sign of the coefficient $\gamma$ when compared to the \textcolor{black}{non-gauge invariant analysis}, the form of the free energy remains the same with the inclusion of vertex corrections, as was anticipated in Ref.~\cite{Setty2020}, and takes the form of leading order $[O(N^0)]$ SYK fluctuation terms~\cite{Stanford2016}. A further Laplace transform yields a fluctuation density of states proportional to $\frac{1}{\omega}$ at low energy. In the \textcolor{black}{non-gauge invariant calculation}, one instead obtains a weaker density of states divergence $\frac{1}{\sqrt{\omega}}$. Finally, vertex corrections render a negative fluctuation correction to the logarithmically divergent term in Eq.~\ref{FP}. Therefore, the curvature of the strong coupling phase diagram is reversed at finite temperatures away from the QCP (see Fig.~\ref{PD}). \par 
 \section{Discussion} Despite the fluctuation free energy acquiring the same form as the corresponding leading order $O(N^0)$ fluctuation contribution of the SYK model, the constant $\gamma$ determining the fluctuation density of states is different in the two cases. This should not come as a surprise since the leading order fluctuation terms for the SYK model are melon diagrams as opposed to pair bubbles that traditionally appear in the theory of fluctuation superconductivity. As emphasized earlier, the self-energy in LS models is proportional to a linear power of the non-interacting Green function (see Eq.~\ref{Repara1}). This is an important property of LSs that is crucial for the symmetry mapping to the $q\rightarrow\infty$ SYK model to work. If the self-energy was, instead, proportional to the total Green function (or the $q$=2 SYK), the model would effectively become the non-interacting random matrix model with no LSs. It is interesting that the form of the gauge-invariant fluctuation propagator near the QCP in Eq.~\ref{FPQCP} has a $\frac{1}{\sqrt{\tau}}$ dependence with the time scale $\tau$ set by $T^{-1}$.  Ref.~\cite{Sachdev2019} used such a form of the conformal propagator and showed that resonant processes produce Planckian scattering rates~\cite{Proust2019, Paglione2019}  with universal coefficients~\cite{Mackenzie2013} independent of interactions. This work, therefore, motivates an evaluation of fluctuation transport quantities such as paraconductivity in the NFL phase close to the QCP using the Larkin-Varlamov formalism~\cite{LarkinVarlamov2005, Setty2019}. It is also interesting to ask whether universality of the butterfly velocity or ``information screening length"~\cite{Baggioli2018} at the QCP holds in the context of Luttinger surfaces.  Finally, in the HK model, the robustness of LSs depends crucially on the ratio of interaction parameter $u$ (or the Mott gap) and bandwidth $W$ of the non-interacting bands~\cite{Phillips2019}. If $W>2 u$, the LS exists only in certain parts of the Brillouin zone.  Same holds true if $2 u>W$ and the system is doped so that the chemical potential is located in one of the Hubbard bands. In either of these cases, one still has extensively many maps to the $q$$\rightarrow \infty$ SYK model -- one for each momentum point where the LS exists. However, the QCP is avoided and the form of the free-energy mapping between the models is lost.\par
   To conclude, we have shown there exists a low-energy reparametrization symmetry in models which host LSs where the self-energy has a simple pole. The transformation properties of the Green function and self-energy can be mapped to the $q$$\rightarrow\infty$ limit of the SYK model. The corresponding mapping of the fluctuation action is robust to inclusion of interaction vertices through the Ward identity, and the subsequent $\frac{1}{\sqrt{\beta u_{c\infty}}}$ behavior of the fluctuation propagator indicates  NFL transport. A LS model of particular interest is the microscopic model by HK~\cite{HK1992}.  In addition to the absence of random interactions, a key simplification of the HK Hamiltonian in Eq.~\ref{kSpaceHK}  is that the interaction terms commute with the kinetic energy making it exactly solvable~\cite{HK1992, Setty2018, Phillips2019}. Despite this simplicity, the model is sufficient to capture important physical phenomena such as LSs, Mott gap and  doublon-holon "Cooper" pairing. Moreover, the model is not restricted to zero dimensions and can be extended to any higher dimensions. Thus the HK model does exactly what is expected of any minimal model -- strip down the full interacting problem to its basic ingredients for describing the most interesting physics. Therefore the problem of pairing instabilities in model LSs, such as those realized in simple microscopic models like HK, lays a firm groundwork toward understanding more sophisticated models exhibiting conformal field theory--gravity duality~\cite{Zaanen2015} and is worth further exploration.   
 \par
 \textit{Acknowledgements:} We thank P. W. Phillips, G. La Nave and L. Yeo for critical comments. This work is supported by the DOE grant number DE-FG02-05ER46236. 
\bibliography{Ward.bib} 
\end{document}